\RequirePackage{fix-cm}
\documentclass[smallextended]{svjour3}       
\smartqed  
\usepackage{graphicx}
\usepackage{amsmath}
\usepackage[T1]{fontenc}
\usepackage{amsmath}
\usepackage[utf8]{inputenc}
\usepackage{mathtools}
\DeclareUnicodeCharacter{2212}{-}
\begin{document}

\title{A geometric look at the objective gravitational  wave function reduction}


\author{Faramarz Rahmani\and Mehdi Golshani\and Ghadir Jafari}
\institute{F. Rahmani \at
Department of Physics, School of Sciences, Ayatollah Boroujerdi University, Boroujerd, Iran\\
              School of Physics, Institute for Research in Fundamental Science(IPM), Tehran, Iran\\
              Tel.: +98-21-22180692,
              Fax: +98-21-22280415\\
              \email{faramarz.rahmani@abru.ac.ir; faramarzrahmani@ipm.ir}           
           \and
           M. Golshani \at
              School of Physics, Institute for Research in Fundamental Science(IPM), Tehran, Iran\\
              Department of Physics, Sharif University of Technology, Tehran, Iran\\
              Tel.:+98-21-66022718, Fax.:+98-21-66022718\\
              \email{golshani@sharif.edu}
              \and
              Gh. Jafari \at
              School of Particles and accelerators, Institute for Research in Fundamental Science(IPM), Tehran, P.O.Box: 19395-5531, Iran\\
ghjafari@ipm.ir
}   

\date{Received: date / Accepted: date}

\maketitle

\begin{abstract}
In ref \cite{RefD1}, a criterion has been derived for the objective wave function reduction through the Shr\"{o}dinger-Newton equation. In this paper, we shall derive that criterion by using the concept of Bohmian trajectories. This study has two consequences. First, providing a geometric perspective on the problem of wave function reduction and the other, representing the role of quantum force and gravitational force in the reduction process.
\end{abstract}

\section{Introduction}
\label{intro}
\begin{figure}[ht] 
\centerline{\includegraphics[width=7cm]{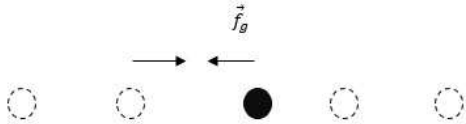}}
\caption{An imagination for the self-gravity of a particle or body in the context of quantum mechanics.The particle has no exact location due to the uncertainty principle.\label{fig:1}}
\end{figure}

One of the questions that has always been raised is the boundary between quantum and classical mechanics. A critical mass may exists for the transition from quantum domain to the classical world. Such critical mass determines the macroscopicity or microscopicity of an object. By knowing the density of matter, the macroscopicty is directly determined in terms of the size of the body.\cite{RefK} From a dynamical point of view, we expect that macroscopic bodies obey the rules of classical mechanics i.e. definite position and momentum, determinism, etc. But, microscopic bodies obey the rules of quantum mechanics, like the uncertainty in position and momentum of the particle. One of the approaches to determine the boundary between quantum mechanics and classical mechanics is gravitational approach. The outstanding  gravitational studies for determining the  boundary between quantum world and the classical world started by Karolyahazy.\cite{RefK}. The remarkable work that was done after that, was Diosi's work, based on the Schr\"{o}dinger-Newton equation. See refs,\cite{RefD1,RefD2,RefD3}. In that equation, there is a term due to the self-gravity of the particle or body. By the self-gravity we mean that according to the Born rule, mass distribution of a particle or body is $\rho = \vert \psi(\mathbf{x},t) \vert ^2$ where can be used to define self-gravity. In other words, we can consider a particle in different locations simultaneously with the distribution $\rho = \vert \psi(\mathbf{x},t) \vert ^2$. Figure (\ref{fig:1}). This is a quantum mechanical concept  which refers to the Heisenberg uncertainty principle and is not obvious in our classical world.
The self gravity of the body or particle would localize the position distribution of the particle or body. It reduces the uncertainty in the position of the particle. \par
Unlike the quantum mechanics, the superposition of different states does not exist in the classical world. A classical system is in a specific state with definite position and momentum. It seems that the classical world has been collapsed by an agent. In the view of some scientists such agent is gravity.\cite{RefD1,RefP1,RefP2,RefP3}
In addition to the gravity, the wave function reduction takes place through a measurement process. In a measurement, the wave function of a quantum system reduces instantaneously to one of the its eigenvectors (wave function collapse). This is one of the postulates of orthodox or standard quantum mechanics, known as collapse postulate. Naturally, there is no room for justifying the collapse phenomena in the standard quantum mechanics. In a measurement process, a pure state which is governed by the linear Schr\"{o}dinger equation, evolves to a mixed state after measurement, through a non-unitary evolution.\cite{RefHolland,RefSak,RefLandsman,RefVon,RefWh,RefBal} This has encouraged some physicists to add so called non-Hamiltonian terms to the Schr\"{o}dinger equation for justifying the non-unitary evolution of the wave function and its reduction.\cite{RefGi1,RefGi2,RefV,RefBassi,RefGisin}.\par In this paper, we do not investigate the problem of the  wave function reduction in the measurement processes. Our aim is the study of gravitational wave function reduction in the Bohmian context. Here, a key question arises. If quantum mechanics is universal, our classical world should be in a superposition. As we mentioned before, our classical world is not in a superposition. For example, a chair in a room is not in its different states of its different degrees of freedom simultaneously. How does our universe collapse to a single state? One of the objective features of bodies which can be considered as a criterion for the distinction between quantum mechanics and classical world is the mass of a body. But, the mass of a body should be considered as an active agent that affects the dynamics of the body. As we mentioned before, one of the famous equations in this regard which gives a criterion for determining a boundary between quantum mechanics and classical mechanics, based on objective properties of matter, is the Schr\"{o}dinger-Newton equation. 
The Schr\"{o}dinger-Newton equation for a single body is:
\begin{equation}\label{sn}
i\hbar \frac{\partial\psi(\mathbf{x},t)}{\partial t}=\left(-\frac{\hbar^2}{2M}\nabla^2 -GM^2 \int \frac{\vert \psi(\mathbf{x}^\prime,t)\vert^2}{\vert \mathbf{x}^\prime -\mathbf{x} \vert} d^3 x^\prime\right) \psi(\mathbf{x},t)
\end{equation}
By using the stationary state $\psi=\psi(\mathbf{x}) e^\frac{iEt}{\hbar}$, for the classical limit, which satisfies the above equation, and using variational method, a relation between the mass of a particle and the width of its associated stationary wave packet is obtained. That relation, which is
\begin{equation}\label{Dio}
\sigma_{0,(min)} = \frac{\hbar^2}{Gm^3}
\end{equation}
provides a criterion  for the transition from the quantum world to the classical world or the breakdown of quantum superposition in terms of universal constants and the mass of the particle. \cite{RefD1}
For a body with radius $R$, the minimum wave packet width is $\sigma^{(R)}=(\sigma_{(min)})^{(\frac{1}{4})}R^{\frac{3}{4}}$. A critical size of a body for transition from quantum domain to classical domain is about $R_c =10^{-5} cm$. For details see \cite{RefK,RefD3}.\par 
After Diosi, the most significant work which has been done, is the gravitational approach of Penrose which is based on two essential concepts in physics: the principle of equivalence and the principle of general covariance.\cite{RefP1,RefP2,RefP3} The overview of Penrose's work is as follow. Consider a body in two different locations with their associated states $\vert \phi_i \rangle, i=1,2$. Each state satisfies the Schr\"{o}dinger equation separately as an stationary state, with a unique Killing vector. The superposed state $\vert \psi \rangle = \alpha \vert \phi_1\rangle +  \beta \vert \phi_2 \rangle $ is also a stationary state with the unique Killing vector $\mathcal{K}=\frac{\partial}{\partial t}$. When, the self gravity of the body(the curvature of the spacetime due to the mass of the object itself ) is considered, the quantum state of the gravitational field of the body at the different locations  i.e, $\vert \mathcal{G}_i \rangle, i=1,2$, must also be taken into account. This changes the state $\vert \psi \rangle$ to the state $\vert\psi_{\mathcal{G}}\rangle = \alpha \vert \phi_1\rangle \vert \mathcal{G}_1 \rangle +  \beta \vert \phi_2\rangle \vert \mathcal{G}_2 \rangle$ which is not a stationary state in the sense that it has not a unique Killing vector. Thus, the total state decays to one of the states to get a stationary state with definite Killing vector of spacetime. In this approach, the decay time for transition from the quantum domain to classical domain is obtained.\cite{RefP1}\par The gravitational considerations of the Penrose which were stated above can be used for justifying the wave function reduction through the measurement process. Usually, in a measurement process we have an apparatus and a microscopic system with their associated wave functions. The quantum state of the apparatus, as a macroscopic body, is entangled with the microscopic or quantum system (an electron for example) during the measurement.\cite{RefHolland,RefP1,RefVon} Since, the apparatus is a macroscopic body, its self-gravity is significant. Then, according to the previous statements, the total entangled state ($\vert \psi_{\mathcal{G}}\rangle$) is not stationary and decays to a specific state for having a definite unique Killing vector. Consequently, the microscopic system  which is entangled with the apparatus also goes to a specific eigenstate. Here, the role of gravity is obvious.\cite{RefP1,RefP2,RefP3}. Why do we want to study this topic in the Bohmian framework?. Because, relativistic Bohmian quantum mechanics can be represented geometrically \cite{RefCarrol,RefShoja,RefFis}. On the other hand, the origin of gravity is the curvature of spacetime which is described geometrically. Thus, we was persuaded to study the objective gravitational wave function reduction in the Bohmian context. \par
\begin{figure}[ht] 
\centerline{\includegraphics[width=6cm]{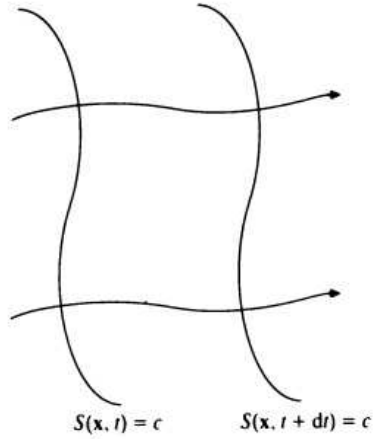}}
\caption{The figure, represents the propagation of the function $S(\mathbf{x},t)$ in the configuration space. Normal to it is the momentum field of the particle.\label{fig:2}}
\end{figure}

 
Bohmian quantum mechanics is a causal and deterministic theory in which  a particle has a definite trajectory with definable physical quantities like in classical mechanics. But it is not a recovery of classical mechanics. Because, the dynamics of the particle is influenced by a pilot-wave or a quantum wave function  $\psi(\mathbf{x},t)$ which has no analogue in classical mechanics. The probability density of \textit{being} a particle in the volume $d^3 \mathbf{x}$ is equal to $\rho(\mathbf{x},t)=R^2(\mathbf{x},t)$. \cite{RefHolland,RefBohm,RefImplicate,Refuniverse}. In Bohm's own view, the departure from classical mechanics appears in an essential entity known as "quantum potential". It is a non-local potential with non-classical features. The primary approach of the Bohm was the substitution of the polar form of the wave function, $\psi(\mathbf{x},t)=R(\mathbf{x},t)\exp(i\frac{S(\mathbf{x},t)}{\hbar})$, into the Schr\"{o}dinger equation which leads to a quantum Hamilton-Jacobi equation as:
\begin{equation}\label{hamilton}
\frac{\partial S(\mathbf{x},t)}{\partial t}+\frac{(\nabla S)^2}{2m}+V(\mathbf{x})+Q(\mathbf{x})=0
\end{equation}
The last term in the above equation is the non-relativistic quantum potential which is given by:
\begin{equation}\label{potential}
Q=-\frac{\hbar^2 \nabla^2 R(\mathbf{x},t)}{2mR(\mathbf{x},t)}=-\frac{\hbar^2 \nabla^2 \sqrt{\rho(\mathbf{x},t)}} {2m\sqrt{\rho(\mathbf{x},t)}}
\end{equation}
The quantum force exerted on the particle is defined as $\mathbf{f}=-\nabla Q$. The function $S(\mathbf{x},t)$, is the action of the system which appears in the phase of the wave function. It is a function in the configuration space of the system. It propagates like a wave front in the configuration space of the system \cite{RefHolland}. Normal to the $S(\mathbf{x},t)$ gives the momentum of the particle through the relation $\mathbf{p}=\nabla S(\mathbf{x},t)$. See figure (\ref{fig:2}). The energy of the particle is obtained using the relation $E=-\frac{\partial S}{\partial t}$.\cite{RefHolland}
With the initial position $\mathbf{x}_0$ and the initial wave function $\psi(\mathbf{x}_0,t_0)$, which gives the initial velocity, the position of the particle is obtained through the guidance equation 
\begin{equation}\label{guidance}
\frac{d\mathbf{x}(t)}{dt}=\left(\frac{\nabla S(\mathbf{x},t)}{m}\right)_{\mathbf{X}=\mathbf{x}(t)}.
\end{equation}
The initial wave function gives the initial phase $S_0$ and consequently initial velocity. With the initial velocity $\mathbf{v}_0 = \frac{\nabla S_0}{m}=\frac{\mathbf{p}_0}{m}$, and the initial position $\mathbf{x}_0$, the evolution of the system is obtained. But always we are faced with a distribution of the initial positions and velocities practically. The dynamics of the system is obtained through the $m\frac{d^2 \mathbf{x}}{dt^2}=\nabla \left(Q(\mathbf{x)+V(\mathbf{x})}\right)$ .\cite{RefHolland}
The expression $\mathbf{X}=\mathbf{x}(t)$ in relation (\ref{guidance}) means that among all possible trajectories, one of them is chosen. Due to our ignorance with respect to the all initial data we are faced with an ensemble of trajectories. The possibility of definition of trajectories in Bohmian quantum physics gives imagination to quantum phenomena and geometric view on quantum mechanics. \par 
In the section (\ref{sec:1}), we shall argue about how it is possible to have an objective explanation for the classical limit of a free particle based on Bohmian trajectories and the self-gravity of the particle. In the section (\ref{sec:2}), we shall derive a Poisson-like relation for the Bohmian quantum potential for the gravitational reduction of the wave function based on deviation of Bohmian trajectories. But for the possibility of developing and generalizing the subject in the future, we do calculations relativisticly, then we shall consider its nonrelativistic limit for deriving Diosi's formula. 
The effect of the gravitational field of the particle is considered as a curved spacetime with a fixed metric tensor $g_{\mu\nu}$. The relativistic generalization of the concept of the self-gravity for a particle in the Schr\"{o}dinger-Newton equation is that the particle is affected by the spacetime curvature due to the its mass-energy. It is natural that if we did not consider the uncertainty in the position of the particle in the framework of the quantum mechanics, such interpretation for self-gravity whether in the relativistic or nonrelativistic domain was not possible.
\section{Toward an objective reduction in Bohmian quantum mechanics}
\label{sec:1}
In the Bohmian quantum mechanics, the usual condition for turning back to classical mechanics is the vanishing of Bohmian quantum potential. The vanishing of the quantum force is needed in some situations.\cite{RefHolland}. These conditions do not give an objective criterion for the reduction of the wave function or the classical limit of a quantum system. In other words, by vanishing the quantum potential or quantum force, one can not estimate the needed mass of an object for transition from quantum domain to classical domain. Its reason is clear. Because, the mass of the particle or body has no an active role in its dynamical evolution. This may achieved by considering the effect of the gravitational force on the particle in its dynamics. The classical limit of a quantum system in Bohmian quantum mechanics has been studied in \cite{RefHolland}.\par Let first study the trajectories of a free particle which is guided by a Gaussian wave packet. The amplitude of the wave packet of a free particle with zero initial group velocity is \cite{RefHolland}:
\begin{equation}\label{rf}
R= (2\pi \sigma^2)^{-\frac{3}{4}} e^{-\frac{\mathbf{x}^2}{4\sigma^2}}
\end{equation}
where, $\sigma$ is the random mean square width of the packet at a time $t$ which represents the spreading of the wave packet.\cite{RefHolland} 
The quantum potential for this system and the quantum force exerted on the particle are obtained through the relations:
\begin{equation}\label{qf}
Q=-\frac{\hbar^2 \nabla^2R}{2mR}= \frac{\hbar^2}{4m\sigma^2}\left(3-\frac{\mathbf{x}^2}{2\sigma^2} \right)
\end{equation}
and
\begin{equation}\label{ff}
f=-\nabla Q = \frac{\hbar^2}{4m\sigma^2}\mathbf{x}
\end{equation}
In Bohmian quantum mechanics the quantum force is responsible for the spreading of wave packet.\cite{RefHolland}. But, in the standard quantum mechanics the dispersion of a wave packet is explained by using the Heisenberg uncertainty principle. 
Now, we want to clarify how the concept of Bohmian trajectories help us to get a criterion for the wave function reduction.\par 
It has been demonstrated in ref \cite{RefHolland} that the trajectory of the particle which is guided by a Gaussian wave packet is:
\begin{equation}\label{tra}
\mathbf{x}(t)=\mathbf{x}_{0}\left(1+(\frac{\hbar t}{2m\sigma_0^2})^2\right)^{\frac{1}{2}}
\end{equation}
where, $\mathbf{x}_{0}$ denotes the initial position of the particle. If the initial position of the particle is at $\mathbf{x}_{0}=0$ (in the middle of the wave packet), the trajectory is a straight line. The quantum force $\mathbf{f}=-\nabla Q$ for such trajectory vanishes and its trajectory is classical. The figure bellow, shows that for initial positions $\mathbf{x}_{0}\neq0$, the trajectories are not straight lines and the particle is affected by a quantum force. In a causal quantum theory, this ensemble refers to existence of the hidden variables.\cite{RefHolland} In standard quantum mechanics this is due to the uncertainty principle.
\begin{figure}[ht] 
\centerline{\includegraphics[width=7cm]{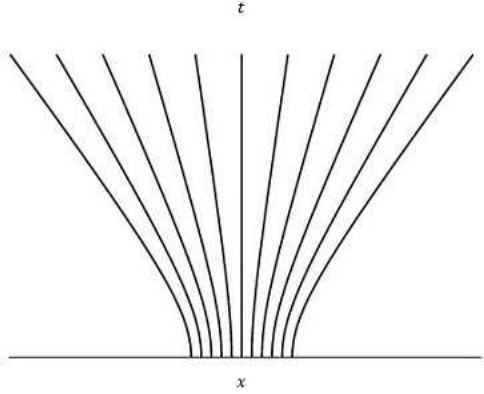}}
\caption{The figure, represents the distribution of trajectories of the particle which is guided by a wave packet in $(1+1)$ dimentional spacetime. Due to uncertainty principle or quantum force, we are faced with an ensemble of trajectories.\label{fig:3}}
\end{figure} 
The figure (\ref{fig:3}) has been depicted using the equation (\ref{tra}). So, up to here, we found that the curved trajectories are the results of the quantum force. For returning them to the classical straight trajectories, we need to an agent for keeping the deviation between trajectories constant through the balancing with the quantum force. The best candidate is the self-gravity of the particle. 
In other words, gravity prevents from more dispersion and consequently more uncertainty in the position of the particle. When the mass of the particle tends to a specified value, its self-gravitation increases. Then, the quantum force and gravitational force can be equal and deviation remains constant. \textit{This is only possible when, the mass of the particle affects its dynamics. Here, the mass of the particle has an active role not a passive one}. Consequently, we should consider the effects of gravity of the particle on its dynamics. In the following, we explain this idea further.\par
\begin{figure}[ht] 
\centerline{\includegraphics[width=7cm]{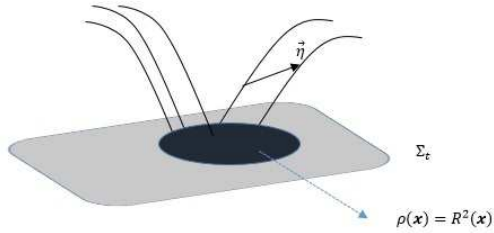}}
\caption{The figure, represents the distribution of timelike trajectories associated with a two-dimensional wave packet on a spacelike hybersurface $\Sigma_t$. Vector $\mathbf{\eta}$, represents the deviation vector between two neighbor trajectories.\label{fig:4}}
\end{figure}

 
Constant deviation has another meaning. That is the width of the wave packet remains constant. In other words we will faced with a wave packet in the form $\psi=R(\mathbf{x}) \exp (-\frac{iEt}{\hbar})$ where, $R(\mathbf{x})$ is the amplitude in the relation (\ref{rf}). 
It may be argued that a particle with a precise location invokes a Dirac delta function, not a stationary wave packet with a constant width. But since the Dirac delta function invokes infinite self-gravity, it is more logical to consider a stationary wave packet rather than a Dirac delta function. The main difference between these two forces is that the gravitational force is defined in a real space, while the quantum force is defined in the configuration space of the particle or system. This type of view, brings up this picture in our mind that in the moment of transition , the quantum effects do not vanish absolutely; rather the self gravity of the particle as an attractive force, does not allow the quantum force to appear.  
When the wave packet evolves according to the Schr\"{o}dinger equation, the particle can be at different locations in the space, with different probability densities $\rho(\mathbf{x})=R^2(\mathbf{x})$. Each position of the particle constructs a trajectory in space-time. By the spreading of the wave packet, the deviation between trajectories changes in time. But, the self-gravity of the particle would converge the trajectories till the the two forces become equal. 
In this regard, we define \textit{a reduced state as a state for which the relative deviation vector between trajectories of the ensemble remains constant}.
The figure (\ref{fig:4}), represents the distribution of an ensemble of timelike trajectories associated with a 2-dimensional wave packet.
\par 
Fortunately, it can be proved that Bohmian trajectories do not cross each other. Thus,
these trajectories can be considered as a congruence. In fact, the single-valuedness of the wave function for a closed loop leads to:
\begin{equation}
\Delta S = \oint_c dS = \oint_c \nabla S \cdot d\mathbf{x}= \oint_c \mathbf{p}\cdot d\mathbf{x}=nh
\end{equation}
Then, the net phase difference around a closed loop is constant and for every instant at the position $\mathbf{x}$, the momentum $\mathbf{p}=\nabla S$ is single-valued.\cite{RefHolland}\par 
In the next section, we demonstrate that  Diosi's formula is obtained through the investigation of the deviation $\eta^{\mu}$ between Bohmian trajectories. 
\section{From the relativistic deviation between Bohmian trajectories to the wave function reduction}
\label{sec:2}
For further developments in the future, we obtain the deviation between trajectories for a relativistic particle. Then, we shall consider its non relativistic limit. But, as we mentioned before, the Bohmian dynamics of the relativistic matter can be described as a  conformal transformation of the background metric. \cite{RefCarrol,RefShoja,RefFis}. Hence, for the richness of the topic we start from the relativistic case. The encounter of fundamental topics like the spacetime geometry and quantum mechanics can be of high value. Karolyahazy in his famous paper \cite{RefK}, started from a relativistic action.  Because he was searching the answer in a more fundamental concept, such as the properties of spacetime. Penrose, too, relates the gravitational collapse to the fundamental properties of the spacetime, like the general covariance and the equivalence principle. \cite{RefP1,RefP2,RefP3}. We shall see that there is a conceptual point in this study. It is the relation between the quantum potential as the manifestation of quntum behavior of matter and the gravity as a property of space time. If, we investigate the problem non-relativistically, we have just done a calculation for obtaining a criterion. But in a relativistic regime,   we have more chance for deepening our physical concepts.\par 
The general scheme is as bellow. Due to the quantum distribution of matter, the definition of self-gravity for a particle is possible. Its relativistic generalization is that the quantum distribution of the particle curves spacetime. Then the dynamics of the particle is affected by this curvature. Also, due to the mass distribution ($\rho =\vert \psi \vert ^2$) we are faced with an ensemble of trajectories (congruence). We represents the metric of such space with the symbol $g_{\mu \nu}$. With this imagination we are going to calculate the deviation between trajectories.\par 
For a spinless relativistic particle, the Hamilton-Jacobi equation and its Bohmian quantum potential are in the form \footnote{The Bohmian quantum potential for the relativistic electron has been derived recently by Hiley and coworkers through complicated methods of Clifford algebra $C^{0}_{3}$.\cite{RefHC} There, the quantum potential is considered as additional term in the Hamilton-Jacobi equation of the electron and is obtained by comparing with the original equation $p_\mu p^\mu =m^2$. In other words, in that case too the relation $p_\mu p^\mu =\mathcal{M}^2$ holds, where, $\mathcal{M}^2=m^2+Q_{Dirac}$.}:
\begin{equation}\label{mass}
 p_\mu  p^\mu =\mathcal{M}^2 
\end{equation}
where
\begin{equation}\label{mq}
\mathcal{M}^2=m^2(1+\mathcal{Q})
\end{equation}
Also,
\begin{equation}
\mathcal{Q}=\frac{\hbar^2}{m^2}\frac{\nabla_\mu \nabla^\mu R}{R}
\end{equation} 
is the relativistic Bohmian quantum potential of a spinless particle in a curved background. \footnote{The substituition of the polar form of the wave function in the Klein-Gordon equation is not the only approach for getting the Hamilton-Jacobi equation (\ref{mass}). For example see the refs \cite{RefCarrol,RefShoja,RefFP,RefRG}.}Here, $R$ stands for the amplitude of the relativistic wave function. \cite{RefHolland,RefCarrol}. The explicit form of $\mathcal{Q}$ is not needed. Because, we need the non-relativistic form of the Bohmian quantum potential (\ref{potential})  for obtaining Diosi's formula. Four-momentum of the particle is defined as:
\begin{equation}\label{fm}
p^\mu =\mathcal{M}u^\mu
\end{equation}
In the presence of the quantum force, the geodesic equation $u^\nu \nabla_\nu u^\mu=0$, with four-velocity $ u^\mu$, is not valid anymore. It will be replaced with the equation
\begin{equation}\label{BG}
 u^\nu \nabla_\nu u^\mu=a^\mu_{(B)}
\end{equation}
where $a^\mu_{(B)}$ refers to the Bohmian acceleration of the particle due to the quantum force. For deriving a relation for the Bohmian acceleration, we start from the relation (\ref{mass}). By differentiation from the both sides of the relation (\ref{mass}) with respect to the parameter $\tau$ which parametrizes the trajectories of the particle, we have:
\begin{equation}\label{m1}
\frac{d}{d\tau}(p_\mu p^\mu)=\frac{d}{d\tau}\mathcal{M}^2
\end{equation}
Then,
\begin{equation}\label{m2}
2p_\mu \frac{dp^\mu}{d\tau}=2\mathcal{M}\frac{d\mathcal{M}}{d\tau}
\end{equation}
 On the other hand, we have:
 \begin{equation}\label{m3}
 \frac{dp^\mu }{d\tau}=\frac{dx^\nu}{d\tau}\nabla_\nu p^\mu = u^\nu \nabla_\nu p^\mu
 \end{equation}
 where we have used the replacement:
 \begin{equation}\label{m4}
 \frac{d }{d\tau} \longrightarrow \frac{dx^\nu}{d\tau}\nabla_\nu = u^\nu \nabla_\nu
 \end{equation}
 Now, by substituting (\ref{m3}) and (\ref{m4}) into the (\ref{m2}), we get:
 \begin{equation}\label{m5}
 p_\mu(u^\nu \nabla_\nu p^\mu)=\mathcal{M} u_\mu \nabla^\mu \mathcal{M}=\mathcal{M} u^\mu \nabla_\mu \mathcal{M}
 \end{equation}
Substituting the relation (\ref{fm}) into the (\ref{m5}), leads to:
\begin{equation}\label{m6}
\mathcal{M}u_\mu (u^\nu \nabla_\nu (\mathcal{M}u^\mu))=\mathcal{M} u_\mu \nabla^\mu \mathcal{M}
\end{equation} 
 Or,
 \begin{equation}\label{m7}
 u^\nu \nabla_\nu (\mathcal{M}u^\mu)=\nabla^\mu \mathcal{M}
 \end{equation}
 which gives:
 \begin{equation}\label{m8}
 u^\mu u^\nu \nabla_\nu \mathcal{M}+\mathcal{M}u^\nu \nabla_\nu u^\mu =\nabla^\mu \mathcal{M}
 \end{equation}
By multiplying both sides of the above equation by $\frac{1}{\mathcal{M}}$, we obtain:
\begin{equation}\label{m9}
u^\nu \nabla_\nu u^\mu = -u^\mu u^\nu \frac{\nabla_\nu \mathcal{M}}{\mathcal{M}}+\frac{\nabla^\mu \mathcal{M}}{\mathcal{M}}
\end{equation}
Now, we should express the right hand side of the above equation in terms of quantum potential. For this purpose, we start from the equation (\ref{mq}) to get:
\begin{equation}\label{m10}
2 \mathcal{M}\nabla_\nu \mathcal{M}=m^2 \nabla_\nu \mathcal{Q} \Rightarrow \frac{\nabla_\nu \mathcal{M}}{\mathcal{M}}=\frac{1}{2}\frac{m^2 \nabla_\nu \mathcal{Q}}{\mathcal{M}^2}
\end{equation}
By using the relation (\ref{mq}) again in the above relation we get:
\begin{equation}\label{m11}
\frac{\nabla_\nu \mathcal{M}}{\mathcal{M}}=\frac{1}{2}\frac{m^2 \nabla_\nu \mathcal{Q}}{m^2(1+\mathcal{Q})}=\frac{1}{2}\frac{\nabla_\nu \mathcal{Q}}{1+\mathcal{Q}}=\frac{1}{2}\nabla_\nu \ln(1+\mathcal{Q})
\end{equation}
Now, we substitute this result into the relation (\ref{m9}) to get:
\begin{equation}\label{m12}
u^\nu \nabla_\nu u^\mu =-\frac{1}{2}u^\mu u^\nu \nabla_\nu \ln(1+\mathcal{Q})+\frac{1}{2}\nabla^\mu \ln(1+\mathcal{Q})
\end{equation}
The left hand side of this equation is the Bohmian acceleration (\ref{BG}). Thus we get Bohmian acceleration in the form:
\begin{equation}\label{ac}
a^\mu_{(B)}=-\frac{1}{2}u^\mu u^\nu \nabla_\nu \ln(1+\mathcal{Q})+\frac{1}{2}\nabla^\mu\ln(1+\mathcal{Q})
\end{equation}
Since the energies, due to quantum potential $\mathcal{Q}$, is very small with respect to the classical energies, we can assume that $\mathcal{Q}<< 1$ and $ln(1+\mathcal{Q}) \simeq \mathcal{Q}$, and the last relation reduces to:
\begin{equation}\label{acac}
a_{(B)}^\mu=-\frac{1}{2}u^\mu u^\nu \nabla_\nu \mathcal{Q}+\frac{1}{2}\nabla^\mu \mathcal{Q}
\end{equation}
This problem is studied in a fixed background metric with the signature $(+1,-1,-1,-1)$ i.e. we do not consider the back-reaction effects of matter on the spacetime.\par
The affine parameter along with a specified trajectory is $\tau$ and tangent to it is the vector $u^\mu=\frac{\partial x^\mu}{\partial \tau}$. The deviation vector between the two neighboring trajectories is defined as $\eta^\mu=\frac{\partial x^\mu }{\partial s}$, in which the parameter $ s$ parameterizes the trajectories so that $\eta^\mu$ is tangent to them. Also, $\eta^\mu$ and $u^\mu$ are orthogonal. The velocity field for the deviation vector  between two neighborhood trajectories is defined as 
\begin{equation}\label{vf}
v^\mu =\frac{d\eta^\mu }{d\tau}=u^\nu \nabla_\nu \eta^\mu
\end{equation}
where we have used from the relation (\ref{m4}).
The acceleration of the deviation vector is:
\begin{equation}\label{bbb}
\frac{dv^\mu }{d\tau}=\frac{d^2 \eta^\mu}{d\tau^2}=\frac{d}{d\tau}(u^\nu \nabla_\nu \eta^\mu)=u^\lambda\nabla_\lambda (u^\nu \nabla_\nu \eta^\mu)
\end{equation}
where we have used the relation (\ref{m4}) again. According to the definitions $\eta^\mu =\frac{\partial x^\mu}{\partial s}$ and $u^\mu = \frac{\partial x^\mu}{\partial \tau}$ and independence of parameters $s$ and $\tau$, we have:
\begin{equation}\label{lie}
\frac{\partial u^\mu}{\partial s}=\frac{\partial}{\partial s}\frac{\partial x^\mu}{\partial \tau}=\frac{\partial}{\partial \tau}\frac{\partial x^\mu}{\partial s}= \frac{\partial \eta^\mu}{\partial \tau}
\end{equation}
But this is the same result that can be reached through the definition of the Lie derivative. In other words,
\begin{equation}\label{lie2}
\frac{\partial u^\mu}{\partial s}=\frac{\partial \eta^\mu}{\partial \tau} \Rightarrow \mathcal{L}_{\mathbf{u}}\eta^\mu = \mathcal{L}_{\mathbf{\eta}}u^\mu =0 \Rightarrow u^\nu \nabla_\nu \eta^\mu = \eta^\nu \nabla_\nu u^\mu
\end{equation}
Now, we substitute this result into the relation (\ref{bbb}) to get:
\begin{equation}\label{ff}
\frac{d^2 \eta^\mu}{d\tau^2}=u^\lambda\nabla_\lambda (\eta^\nu \nabla_\nu u^\mu)
\end{equation}
In fact, substituting the relation (\ref{lie2}) into the relation (\ref{bbb}), gives the deviation acceleration in terms of velocity field derivative $\nabla_\nu u^\mu$.
Now, we expand the relation (\ref{ff}). 
\begin{equation}\label{fff}
\frac{d^2 \eta^\mu}{d\tau^2}=(u^\lambda \nabla_\lambda \eta^\nu)\nabla_\nu u^\mu + u^\lambda \eta^\nu (\nabla_\lambda \nabla_\nu u^\mu)
\end{equation}
For the first term of the above equation, we use the result (\ref{lie2}). This gives:
\begin{equation}\label{n1}
\frac{d^2 \eta^\mu}{d\tau^2}=(\eta^\lambda \nabla_\lambda u^\nu)\nabla_\nu u^\mu + u^\lambda \eta^\nu (\nabla_\lambda \nabla_\nu u^\mu)
\end{equation}
For the second term, we use the curvature formula $\left[\nabla_\lambda \nabla_\nu - \nabla_\nu \nabla_\lambda \right]A^\mu =R^\mu_{\rho\lambda\nu}A^\rho $. Here, $R^\mu_{\rho\lambda \nu}$ denotes the curvature due to the mass-energy of the particle. For the field, $u^\mu$, which is tangent vector to the trajectories $x^\mu(\tau)$, we have:
\begin{equation}\label{n2}
\left[\nabla_\lambda \nabla_\nu - \nabla_\nu \nabla_\lambda \right]u^\mu =R^\mu_{\rho\lambda\nu}u^\rho
\end{equation}
Or,
\begin{equation}\label{n3}
\nabla_\lambda \nabla_\nu u^\mu =\nabla_\nu \nabla_\lambda u^\mu +R^\mu_{\rho\lambda\nu}u^\rho
\end{equation}
By substituting this result into the second term of (\ref{n1}) we have:
\begin{eqnarray}\label{n4}
\frac{d^2 \eta^\mu}{d\tau^2}&=&(\eta^\lambda \nabla_\lambda u^\nu)\nabla_\nu u^\mu+ u^\lambda \eta^\nu(\nabla_\nu \nabla_\lambda u^\mu +R^\mu_{\rho\lambda\nu}u^\rho)\nonumber \\
&=& (\eta^\lambda \nabla_\lambda u^\nu)\nabla_\nu u^\mu + u^\lambda \eta^\nu(\nabla_\nu \nabla_\lambda u^\mu )+R^\mu_{\rho\lambda\nu}u^\rho u^\lambda \eta^\nu
\end{eqnarray}
The second term of the above equation is equal to:
\begin{equation}\label{n5}
u^\lambda \eta^\nu(\nabla_\nu \nabla_\lambda u^\mu) =\eta^\nu \nabla_\nu (u^\lambda \nabla_\lambda u^\mu)-(\eta^\nu \nabla_\nu u^\lambda)(\nabla_\lambda u^\mu)
\end{equation}
The second term of this result cancels out the first term  of (\ref{n4}), after substitution (\ref{n5}) into the (\ref{n4}). Note that $\nu$ and $\lambda$ are dummy indices and the replacement $\nu \longleftrightarrow \lambda$ is allowed. Thus, relation (\ref{n4}) reduces to the:
\begin{equation}\label{n6}
\frac{d^2 \eta^\mu}{d\tau^2}=\eta^\lambda \nabla_\lambda  (u^\nu \nabla_\nu u^\mu)+R^\mu_{\rho\lambda\nu}u^\rho u^\lambda \eta^\nu
\end{equation}
In the presence of gravity alone, the first term vanishes, because the expression in the parentheses satisfies geodesics equation. But here, a quantum force exists. Thus the expression in the parentheses is equal to equation (\ref{BG}). 
Then, we have:
\begin{equation}\label{aa3}
\frac{d^2 \eta^\mu}{d\tau^2}=\eta^\lambda \nabla_\lambda a^\mu _{(B)} + R^\mu _{\rho \lambda \nu}u^\rho  \eta ^\lambda u^\nu .
\end{equation}
The relation (\ref{aa3})  can be expressed in terms of quantum potential: 
\begin{equation}\label{12}
\frac{d^2 \eta^\mu}{d\tau^2}=\eta^\lambda \nabla_\lambda\left(-\frac{1}{2}u^\mu u^\nu \nabla_\nu \mathcal{Q} +\frac{1}{2}\nabla^\mu \mathcal{Q}\right)+R^\mu _{\rho \lambda \nu}u^\rho  \eta ^\lambda u^\nu 
\end{equation}
Where, we have used the relation (\ref{acac}). Now, we need its nonrelativistic limit to demonstrate the correctness of our previous arguments about the role of quantum and gravitational forces in the objective gravitational collapse of the wave function.\par
At the non-relativistic limit, we can take $u^i \simeq \delta^\mu_0$, $\tau \rightarrow t$, $\nabla_\mu \rightarrow \partial_\mu$ and $R^\mu_{0\rho 0}=\partial^\mu \partial_\rho \varphi(x)$, where $\varphi(x)$ is the Newtonian gravitational potential. Also the non-relativistic Bohmian quantum potential has no explicit dependence on time. In other words, $\partial_0 Q=0$. (See relation (\ref{potential})). Then, the equation (\ref{12}) takes the form:
\begin{equation}\label{nredev}
\frac{\partial^2 \eta^i}{\partial t^2}=\eta^j \partial_j \left(\frac{\partial^i Q}{m} - \partial^i \varphi(\mathbf{x}) \right)
\end{equation}
One of the possibilities for having constant deviation or parallel trajectories in the ensemble in the non-relativistic domain, is that in the above equation we take:
\begin{equation}\label{pc}
\partial^i Q = m \partial^i \varphi(\mathbf{x}),\quad i=1,2,3 \quad \text{or} \quad \nabla Q =m\nabla \varphi
\end{equation}
This relation represents the equivalence of Bohmian quantum force and self-gravitational force of the particle for constant deviation between trajectories. Thus, the validity of our physical argument i.e. the equivalence of self-gravitational force with the quantum force in the transition regime is confirmed here. For obtaining an objective criterion for the transition from the quantum domain to the classical world, we do as bellow. For simplicity, we do calculations for a one-dimensional stationary wave packet, with the width $\sigma_0$. 
If we calculate the average quantum potential for a stationary one-dimensional wave packet $\psi(x,t)= (2\pi \sigma_0^2)^{-\frac{1}{4}}e^{-\frac{x^2}{4\sigma_0^2}} e^{\frac{iEt}{\hbar}}$,  with $R_s(x)=(2\pi \sigma_0^2)^{-\frac{1}{4}}e^{-\frac{x^2}{4\sigma_0^2}} $, we get:
\begin{equation}\label{aq}
\langle Q\rangle _{s}=\int_{-\infty}^{+\infty} R_s^2  Q_s  dx  =\int_{-\infty}^{+\infty} R_s^2  \left(-\frac{\hbar^2}{2m}\frac{\nabla^2 R_s}{R_s}\right)  dx \sim \frac{\hbar^2}{2m\sigma_0^2}
\end{equation}
which is the average quantum potential of the particle, when it is described by a stationary wave packet with the width  $\sigma_0$. In general the evolution of the wave packet is not stationary. In Bohmian quantum mechanics, for a stationary wave packet we have $\mathbf{p}=\nabla S(t)=0$, because for the stationary wave functions, the phase of the wave, is a function of time only. Therefore, the kinetic energy of the particle vanishes, and the energy of the particle is due to the quantum potential completely. While, in standard quantum mechanics it is due to kinetic term $\frac{\mathbf{p}^2}{2m}$. The result of (\ref{aq}), i.e $\frac{\hbar^2}{2m\sigma_0^2}$, is exactly same as kinetic term in standard quantum mechanics. There, it is obtained through the relation 
\begin{equation}
\langle K\rangle _{s}=\int_{-\infty}^{+\infty} \psi^{*}_s\left(\frac{\hat{\mathbf{p}}^2}{2m}\right)\psi_s dx \sim \frac{\hbar^2}{2m\sigma_0^2}
\end{equation}
This result has been obtained in ref \cite{RefD1}. \par 
The gravitational self energy for a particle with mass $m$ is defined as:
\begin{equation}\label{self}
U_{g}(\mathbf{x})=-Gm^2 \int \frac{\vert \psi(\mathbf{x}^\prime,t)\vert^2}{\vert \mathbf{x}^\prime -\mathbf{x} \vert} d^3 x^\prime
\end{equation}
where, $\vert \psi(\mathbf{x}^\prime,t)\vert^2 = R^2_s(\mathbf{x}^\prime)$.
The average gravitational self energy of the particle in one dimension, with the width $\sigma_0$, is: 
\begin{equation}\label{ag}
\langle U _{g}\rangle = \int_{-\infty}^{+\infty} R_s^2 U _{g} dx =-Gm^2\int_{-\infty}^{+\infty} \int_{-\infty}^{+\infty} \frac{R^2_s(x^\prime)R^2_s(x)}{\vert x^\prime -x \vert} dx  dx^\prime\sim -\frac{Gm^2}{\sigma_0}
\end{equation}
It is obvious from the relations (\ref{self})and(\ref{ag}) that the average quantum potential and average self-gravitational energy are functions of $\sigma_0$. Because we are actually dealing with average values, we obtain the critical width for which the average gravitational force and quantum force become equal:
\begin{equation}\label{dio}
\frac{d\langle Q\rangle _{s}}{d\sigma_0} =\frac{d\langle U _{g}\rangle}{d\sigma_0}  \quad \Rightarrow \quad (\sigma_0)_{\text{critical}}\sim \frac{\hbar^2}{Gm^3}
\end{equation}
Where, derivative has been taken with respect to the probabilistic width $\sigma_0$. 
\textit{The above relation is the famous result of Diosi that has been obtained here from a geometrical approach ralated to Bohmian trajectories}. See ref \cite{RefD1}. This topic, reveals the  role of gravity in the wave function reduction in the context of Bohmian quantum mechanics. \cite{RefK,RefBassi}. \par 
By imposing the operator ($\nabla \cdot $) on the both sides of the condition (\ref{pc}), we get the a Poisson-like equation for the quantum potential as:
\begin{equation}\label{prf}
\nabla^2 Q = 4\pi G m \rho \quad \text{or} \quad -\frac{\hbar^2}{2m^2}\nabla^2(\frac{\nabla^2 \sqrt{\rho}}{\sqrt{\rho}})=4\pi G \rho
\end{equation}
This is a non linear differential equation which expresses that, in the moment of transition from quantum world to the classical world, quantum information ($Q$) is replaced by gravitational information ($\varphi$). 
The equation (\ref{prf}) can be solved for different distributions $\rho$ to get more insight to the problem. We have derived this relation in the ref \cite{RefRGG} from another point of view. Then we investigated the behavior of $\rho$ in the wave function reduction with respect to the mass variations. Relation (\ref{prf}) is the consequence of the gravitational wave function reduction in Bohm's causal quantum theory. This may be a starting point for further studies in this context.
\section{conclusion}
\label{sec:3}
In this study, we argued how it is possible to obtain a criterion for the gravitational objective wave function reduction in Bohmian quantum mechanics related to the quantum potential or quantum force. It was done based on Bohmian trajectories and geometrical concepts. Finally, in addition to the result of the previous work of Diosi, an interesting nonlinear equation, i.e. equation (\ref{prf}), was obtained. It represents a Piosson-like equation for the Bohmian quantum potential in the moment of reduction. Its solutions can be investigated numerically or analytically. In fact, we have stated that quantum information reduces to the gravitational information at the reduction time, where, expressed as a balance between self-gravitational force of the particle and its quantum force. In Bohmian quantum mechanics, quantum potential is responsible for the quantum behavior of a system. On the other hand, according to the role of quantum potential and gravity in the reduction process, we guess that such studies  may be a clue for investigating the possible connections between quantum mechanics and gravity; at least in the Bohmian context. In the spirit of Bohmian quantum mechanics as a causal quantum theory, the existence of some hidden variables causes the quantum behavior of matter. Unknowing such variables leads to probabilistic interpretation for quantum mechanics. Obviously, this is contrary to the fundamentals of standard quantum mechanics in which the probabilistic interpretation is an intrinsic aspect of the universe. In general,in Bohmian quantum mechanics the quantum force is responsible for the dynamical evolution of matter. So, the quantum force should be related to the such hidden variables. In transition from the quantum domain to the classical world, the quantum force becomes equal to the gravitational force. It seems that the clue of such hidden variables may be found between the gravitational concepts.

\end{document}